\documentclass[aps,prd,superscriptaddress,twocolumn]{revtex4}
\usepackage[normalem]{ulem}
\usepackage{epsfig}
\usepackage{amsfonts}
\usepackage{amsmath}
\usepackage{slashed}
\usepackage{graphicx}
\usepackage{color}
\usepackage{mathtools}
\usepackage{extarrows}
\allowdisplaybreaks[4]

\begin{document}
	
	\preprint{JLAB-THY-19-3065}
	
	\title{Heavy quark expansion for heavy-light light-cone operators }

	\author{Shuai Zhao}
	\email{szhao@odu.edu}
	\affiliation{Department of Physics, Old Dominion University, Norfolk, VA 23529, USA}
	\affiliation{Theory Center, Thomas Jefferson National Accelerator Facility, Newport News, VA 23606, USA}
	
	\begin{abstract} 
		 We generalize the celebrated heavy quark expansion to nonlocal QCD operators. By taking nonlocal heavy-light current on the light-cone as an example, we confirm that the collinear singularities are common between QCD operator and the corresponding operator in heavy quark effective theory (HQET), at the leading power of $1/M$ expansion. Based on a perturbative calculation in operator form at one-loop level, a factorization formula  linking QCD and HQET operators is investigated and the matching coefficient is determined. The matching between QCD and HQET light-cone distribution amplitudes (LCDAs) as well as other momentum distributions of hadron can be derived as a consequence.
	\end{abstract}


	\maketitle


Hadrons are multi-scale strong interaction systems. Heavy hadron---the hydrogen atom of strong interaction, plays an unique role of understanding and examining quantum chromodynamics (QCD).  When one of the quarks in a hadron is heavy comparing with strong interaction scale, i.e., $M\gg \Lambda_{\mathrm{QCD}}$, the hard scale $M$ is expected to disentangle from the infrared scale.  This leads to the heavy quark effective theory (HQET)~\cite{Georgi:1990um,Eichten:1989zv,Isgur:1989vq}, which has proved an effective approach of studying heavy flavor hadrons, especially in $B$-meson physics. For a review of HQET, see Refs.~\cite{Neubert:1993mb,Manohar:2000dt}. 

The HQET action can be derived by expanding the QCD action in series of the inverse powers of $M$, which is known as the heavy quark expansion (HQE). The HQE for local composite operators is also extensively explored. For example, consider the heavy-light axial-vector current $\bar q\gamma^{\mu} \gamma^5 Q$, its HQE gives
\begin{align}
\bar q\gamma^{\mu}\gamma^5 Q = C(M,\mu) \bar q\gamma^{\mu}\gamma^5 h_v +\mathcal{O}(1/M),\label{eq:HQE:local}
\end{align}
where $\bar q$ is light quark, $Q$ is the heavy quark field in QCD, while $h_v$ is the heavy quark field in HQET, with velocity index $v$. A  matching coefficient $C(M,\mu)$ is introduced due to the different ultraviolet (UV) behavior of the full and effective theories. The matching coefficient can be calculated in perturbation theory, while the infrared physics is only enclosed in the operators. This relation holds at operator level, so the matching equation as well as the matching coefficient are independent of hadron states.

Even in local field theories, one can construct not only local composite operators, but also nonlocal operators. In QCD and its effective theories, the nonlocal operators are crucial for understanding inner structure of hadrons. One important type of such operators are the bilocal quark operators $\bar q(z)[z,0]\Gamma Q(0)$, in which the two quark fields are located on the light-cone (i.e., $z^2=0$ but $z\neq 0$), with $\mu$ being the renormalization scale that defines the operator. The parton momentum distributions  in a hadron, e.g., parton distribution functions (PDFs) and light-cone distribution amplitudes (LCDAs), are defined through the matrix elements of light-cone operators. These distributions are indispensable ingredients for QCD factorization theorems. For example, for many $B$-meson exclusive decay processes, the decay amplitude can be factorized in terms of hard scattering kernel and $B$-meson LCDAs~\cite{Beneke:1999br,Beneke:2000wa,Beneke:2001at,Bosch:2001gv,Becher:2005fg,Lu:2018cfc,Gao:2019lta}, where the $B$-meson LCDAs are defined by the matrix elements of heavy-light operators on the light-cone in HQET~\cite{Grozin:1996pq}.  The other case  is that the two parton fields are separated off the light-cone. The space-like operators attract lots of attentions in the past few years, thanks to the development of large momentum effective theory~\cite{Ji:2013dva,Ji:2014gla} and many other approaches designed for accessing parton physics from lattice calculation, e.g., pseudo-PDFs~\cite{Radyushkin:2017cyf,Orginos:2017kos} and lattice cross-sections~\cite{Ma:2014jla,Ma:2017pxb}.

When the heavy quark mass $M\gg \Lambda_{\mathrm{QCD}}$, analogous to local operators, the bilocal operators are also expected to be factorized into hard functions and HQET bilocal operators. The matching for the first inverse moment of LCDAs in QCD and HQET was derived in Ref.~\cite{Pilipp:2007sb}. A factorization theorem for LCDAs was proposed recently~\cite{Ishaq:2019dst}, which connects $B$-meson LCDAs defined in QCD and HQET, based on the perturbative calculation on the LCDAs of heavy-light mesons~\cite{Bell:2008er}. In this work, we will focus on the operators instead of the momentum distributions, because factorization holds at operator level, taking matrix elements and Fourier transforms are irrelevant for establishing a factorization theorem. 

The goal of this work is to derive the HQE for nonlocal QCD operators, or in other words, the nonlocal generalization of Eq.~\eqref{eq:HQE:local}. 
Without loss of generality,
we will study the HQE for the nonlocal heavy-light current in which two quark fields are separated on the light-cone, similar discussions might be easily generalized to other nonlocal operators. 
Based on the factorization formula in the operator form, the factorization for $B$-meson LCDAs and other structure functions can be naturally derived.


In QCD, a gauge invariant nonlocal light-cone operator composed by a light quark field and a heavy quark field can be expressed as $\bar q(z)[z,0]\Gamma Q(0)$,
where $\bar q(z)$ denotes the light-quark with mass $m$, and $M$ is the mass of the heavy quark field $Q(0)$, $[z,0]\equiv P\exp[ig_s\int_0^1 d\lambda z\cdot A(\lambda z)]$ is a Wilson line located on the light-cone. The position of light quark is $z=z^- n$, with $n$ being a unit light-cone vector, $n^2=0$. $\Gamma$ is a certain Lorentz structure. For the sake of simplicity, we consider a special case $O(z,0)\equiv \bar q(z)[z,0]\slashed n \gamma^5 Q(0)$, which corresponds to the leading twist LCDA of $B$-meson. The corresponding HQET operator is denoted as  $\widetilde{O}(z,0;v)\equiv\bar q(z)[z,0]\slashed n\gamma^5 h_v(0)$, where $h_v$ is the heavy quark field in HQET, related to the large component of $Q$ under $M\to\infty$ limit, $v$ ($v^2=1$) is the velocity vector of heavy quark.  $h_v$ is constrained by $\slashed v h_v=h_v$ and equation of motion $v\cdot D~ h_v=0$.

When the heavy quark mass $M$ is large, the heavy quark and QCD Lagrangian can be expanded in series of $1/M$. The full heavy quark field $Q$ is expressed by the effective heavy quark field $h_v$ as (see, e.g., Refs.~\cite{Neubert:1993mb,Manohar:2000dt})
\begin{align}
 Q(x) &=e^{-i M v \cdot x}\left(1+\frac{i \slashed  D_{\perp}}{2 M}+\ldots\right) h_{v}(x) , \label{eq:exp:field}
\end{align}
where $D_{\perp}^{\mu}\equiv D^{\mu}-v^{\mu} v\cdot D$, with $D$ denoting the covariant derivative. 
At tree level, since there is no interaction, we immediately have
\begin{align}
O(z,0)^{(0)}=\widetilde{O}(z,0;v)^{(0)}+\mathcal{O}\left(\frac{1}{M}\right),
\end{align}
with the help of Eq.~\eqref{eq:exp:field}.
The operators with superscript $(0)$ denotes the un-corrected operators.

If the radiative corrections are included, however, HQE will generally modify the UV behavior. Taking $M\to \infty$ in radiative correction of $O(z,0)$ cannot be reduced to $O(z,0;v)$ when ultraviolet (UV) singularities exist,  then a matching is needed. Because the matching is related to hard scale $M$, the matching coefficient can be evaluated in perturbation theory.
When the interaction is included, the position of quarks will be generally  shifted, which means that HQE of $O(z,0)$ will be a superposition of $O(\bar \alpha z,\beta z ;v)$, with $0<\beta<\bar \alpha<1$. The HQE formula proposed in this work is
\begin{align}
O(z,0,\mu)
=&\int_0^1 d\alpha \int_0^{\bar \alpha} d\beta~ C(\alpha,\beta,t,M,\mu,\widetilde{\mu};\alpha_s)\nonumber\\
&\cdot\widetilde{O}(\bar\alpha z,\beta z,\widetilde{\mu};v)+\mathcal{O}\left(\frac{1}{M}\right),\label{eq:fac}
\end{align}
where $t \equiv v\cdot z-i 0$, $\bar\alpha\equiv 1-\alpha$. $C(\alpha,\beta, t ,M,\mu,\widetilde{\mu};\alpha_s)$ is the matching coefficient, which can be evaluated in perturbation theory.
To confirm the matching formula and evaluate the matching coefficient, one should first calculate the radiative corrections of both QCD and HQET operators.  


Since the nonlocal operator is defined in position space, it is natural to perform calculation in coordinate-representation. Furthermore, the coordinate-representation calculation can be done in operator form.   We work in $D=4-2\epsilon$ dimensions so that the UV and soft singularities are regularized in dimensional regularization (DR). The light-quark mass $m$ serves as the regulator for the collinear (mass) singularity.

The radiative corrections to operator $O(z,0)$ involve UV singularity, so the operator should be renormalized first. Here we adopt the modified minimal subtraction ($\overline{\mathrm{MS}}$) scheme. 
 The renormalization group equation (RGE) for $O(z,0,\mu)$ is~\cite{Balitsky:1987bk}
\begin{align}
\mu^2\frac{d}{d\mu^2}O(z,0,\mu)=\int_0^1 d\alpha\int_0^{\bar \alpha}d\beta~ V(\alpha,\beta) O(\bar\alpha z,\beta z,\mu),
\end{align}
and 
\begin{align}
V(\alpha,\beta)=&\frac{\alpha_s C_F}{2\pi}\bigg(\delta(\beta)\left[\frac{\bar \alpha}{\alpha}\right]_+ + \delta(\alpha)\left[\frac{\bar \beta}{\beta}\right]_+ \nonumber\\
&+ 1-\frac12\delta(\alpha)\delta(\beta)\bigg)+\mathcal{O}(\alpha_s^2)
\end{align}
is the Balitsky-Braun evolution kernel, where the plus distribution is defined by
\begin{align}
\int_0^1 du \bigg[\frac{\bar u}{u}\bigg]_+ T(u)\equiv \int_0^1 du \frac{\bar u}{u} [T(u)-T(0)],\label{eq:plus}
\end{align}
with $T(u)$ denoting a test function.
 It indicates that under renormalization, the nonlocal operator will get mixed with all the operators of the same type but with smaller separation between two quarks.
By taking the forward hadron-to-hadron or meson-to-vacuum matrix elements and performing Fourier transform, this equation will be reduced to the nonsinglet part of the Dokshizer-Gribov-Lipatov-Altarelli-Parisi equation for PDFs~\cite{Altarelli:1977zs,Dokshitzer:1977sg,Gribov:1972ri}, or the Efremov-Radyushkin-Brodsky-Lepage equation for LCDAs ~\cite{Lepage:1979zb,Efremov:1978rn,Efremov:1979qk}, respectively~\cite{Mueller:1998fv}. Recently the evolution of light-cone operators are known up to three-loops~\cite{Braun:2014vba,Braun:2016qlg,Braun:2017cih,Braun:2019qtp}.

The renormalized operators including radiative correction can be generally expressed as
\begin{align}
O(z,0,&\mu)^{\mathrm{ren.}}=\int_0^1 d\alpha \int_0^{\bar\alpha} d\beta~ K(\alpha,\beta, m, M,\mu;\alpha_s) \nonumber\\
&\cdot O(\bar \alpha z,\beta z)^{(0)}+\mathrm{higher~ twist~ operators}.\label{eq:exp}
\end{align}
 Here the operators that vanished by equation of motion are also eliminated.
The function $K(\alpha,\beta, m, M,\mu;\alpha_s)$ is a series in $\alpha_s$
\begin{align}
K(\alpha,&\beta, m, M,\mu;\alpha_s)=K^{(0)}(\alpha,\beta)\nonumber\\
&+\frac{\alpha_s C_F}{2\pi}K^{(1)}(\alpha,\beta, m, M,\mu)+\mathcal{O}(\alpha_s^2), 
\end{align}
with  $K^{(0)}(\alpha,\beta)=\delta(\alpha)\delta(\beta)$. The one-loop term can be calculated in coordinate-representation. The result reads
\begin{widetext}
\begin{align}
K^{(1)}(\alpha, \beta, & m, M,\mu)=\delta Z~\delta(\alpha)\delta(\beta)+\bigg[\frac{\bar \alpha}{\alpha}\ln\frac{\mu^2}{\alpha^2  u_0^2 M_H^2}\bigg]_+\delta(\beta)+\bigg[\frac{\bar \beta}{\beta}\ln\frac{\mu^2}{\beta^2 \bar u_0^2 M_H^2}\bigg]_+\delta(\alpha)\nonumber\\
&+\frac{2 u_0 \bar u_0+(\alpha u_0- \beta \bar u_0)(u_0-\bar u_0)-(\alpha u_0-\beta \bar u_0)^2}{[(\alpha u_0-\beta \bar u_0)^2]^{1+\epsilon_{\mathrm{IR}}}}\Gamma(1+\epsilon_{\mathrm{IR}})\bigg(\frac{\mu^2_{\mathrm{IR}} e^{\gamma_E}}{M_H^2}\bigg)^{\epsilon_{\mathrm{IR}}}+\ln\frac{\mu^2}{M_H^2(\alpha u_0-\beta\bar u_0)^2},
\label{eq:coordinate}
\end{align}
\end{widetext}
where $M_H\equiv m+M$, and $u_0\equiv m/M_H$, $\mu$ and $\mu_{\mathrm{IR}}$ are the renormalization and soft scales, respectively, $\gamma_E$ is the Euler–Mascheroni constant,  $(\alpha_s C_F/2\pi)\delta Z=\sqrt{Z_{2,q}^{\mathrm{OS}}Z_{2,Q}^{\mathrm{OS}}}-1$,  
$Z_{2,q}^{\mathrm{OS}}$ and $Z_{2,Q}^{\mathrm{OS}}$
are the $\mathrm{\overline{MS}}$ subtracted on-shell renormalization constants for $q$ and $Q$, respectively. The second term in Eq.~\eqref{eq:coordinate} is from the interaction between light quark and Wilson line, while the third term is from heavy quark---Wilson line interaction. The last two terms are from light quark---heavy quark interaction. Note that there is a scheme dependence on the treatment of $\gamma^5$ in DR: one is the naive DR scheme that $\gamma^5$  anti-commutes with all $\gamma^{\mu}$~\cite{Chanowitz:1979zu}; another choice is the 't Hooft-Veltman scheme~\cite{tHooft:1972tcz,Breitenlohner:1977hr}, in which $\gamma^5$ anti-commutes with $\gamma^{\mu}$ for $\mu=0,1,2,3$ but commutes with $\gamma^{\mu}$ for $\mu=4,\cdots,d-1$. Without loss of generality, we simply adopt naive scheme in this work. 
We also note that  $\epsilon_{\mathrm{IR}}$ is not expanded at this stage, because 
the existence of soft singularities located at $\alpha=\beta=0$. Such expansion is only safe when the soft singularities are isolated (e.g., by introducing plus-prescriptions for the integrals). 

Our result in Eq.~\eqref{eq:coordinate} is valid for arbitrary $m$ and $M$. To compare with previous result on LCDA for mesons with non-equal quark masses (e.g., $K$ and $B_c$), one can sandwich the operator $O(z,0,\mu)$ between vacuum and the lowest Fock state, then Fourier transform to momentum space. By recalling Eq.~\eqref{eq:exp}, this is equivalent to a convolution between $K$ and $\delta(x-\bar\alpha u_0-\beta\bar u_0)$.
 With the kernel given in Eq.~\eqref{eq:plus}, and eliminating the contribution from decay constant,
we will arrive at the result for LCDA, which was firstly calculated by Bell and Feldmann~\cite{Bell:2008er} and later further explored in NRQCD re-factorization approach~\cite{Xu:2016dgp,Wang:2017bgv}.
Another special case is $u_0=1/2$, i.e., $m=M$, then Eq.~\eqref{eq:coordinate} describes the one-loop correction to the operator with equal quark masses, which can be used to mesons like $\pi^0$ and $\eta_c$, etc.

Since the topic of this work is the matching of heavy-light operator, what we are interested in is the $M\to \infty$ limit. After some efforts, we arrive at
\begin{align}
&~~K^{(1)}(\alpha, \beta, m, M,\mu)\nonumber\\
=&\left(\frac34\ln\frac{M^2}{m^2}-3\right)\delta(\alpha)\delta(\beta)+\bigg[\frac{\bar \beta}{\beta}\ln\frac{\mu^2}{\beta^2 M^2}\bigg]_+\delta(\alpha)\nonumber\\
&+\bigg[\frac{\bar \alpha}{\alpha}\ln\frac{\mu^2}{\alpha^2 m^2}+\frac12\ln\frac{\bar \alpha^2\mu^2}{\alpha^2 m^2}-\frac{2}{\alpha}+\frac32\bigg]_+\delta(\beta)\nonumber\\
&+\bigg[\frac{\bar \beta}{\beta}+\ln\frac{\mu^2}{\beta^2 M^2}\bigg]_+ +\mathcal{O}\left(\frac{1}{M}\right).\label{eq:kernel:qcd}
\end{align}



The one-loop correction to HQET operator can be calculated in the same manner with QCD case. We denote HQET operator as $\widetilde{O}(z,0;v)\equiv \bar q(z)[z,0]\slashed n\gamma_5 h_v(0)$, and add a tilde upon other related variables  to distinguish from the QCD ones. We adopt the $\overline{\mathrm{MS}}$ scheme again for renormalization.
Unlike the QCD case, there is a $1/\epsilon_{\mathrm{UV}}^2$  UV divergence. In HQET, the heavy quark is described by a Wilson line along the $v$-direction. The interaction between the $v$- and $n$-Wilson lines  generates a cusp singularity, therefore light-cone singularity and cusp singularity appear simultaneously and leads to the $1/\epsilon_{\mathrm{UV}}^2$-pole. The cusp singularity and corresponding cusp anomalous dimension was computed at two-loop order long time ago~\cite{Korchemsky:1987wg,Korchemskaya:1992je} and recently has been known up to three-loops~\cite{Grozin:2014hna,Grozin:2015kna}. The light quark---Wilson line interaction contributes equally to both QCD and HQET operators. The heavy quark---light quark interaction is UV finite. The RGE for $\widetilde{O}(z,0,\widetilde{\mu};v)$ is
 \begin{align}
&\widetilde{\mu}^2\frac{d}{d\widetilde{\mu}^2}\widetilde{O}(z,0,\widetilde{\mu};v)\nonumber\\
=&-\frac{\alpha_s C_F}{2\pi}\left[ \ln(i t e^{\gamma_E}\widetilde{\mu} )-\frac14\right]\widetilde{O}(z,0,\widetilde{\mu};v)\nonumber\\
&+\frac{\alpha_s C_F}{2\pi}\int_0^1 d\alpha\left[\frac{\bar\alpha}{\alpha}\right]_+\widetilde{O}(\bar\alpha z,0,\widetilde{\mu};v).
\end{align}
If the anomalous dimension from decay constant is counted, this evolution equation will match the RGE for $B$-meson LCDA in coordinate space~\cite{Kawamura:2010tj}.  The RGE for $B$-meson LCDA in the name of Lange-Neubert equation was first derived in momentum space~\cite{Lange:2003ff}. The two-loop evolution equation was derived very recently~\cite{Braun:2019wyx}.

 After the UV singularities are removed, the renormalized HQET operator is linked to the tree-level one by
\begin{align}
\widetilde{O}(z,0,&\widetilde{\mu}; v)^{\mathrm{ren.}}=\int_0^1 d\alpha \int_0^{\bar\alpha} d\beta~ \widetilde{K}(\alpha,\beta, m, t, \widetilde{\mu};\alpha_s)\nonumber\\
&\cdot \widetilde{O}(\bar \alpha z,\beta z;v)^{(0)}+\mathrm{higher~ twist~ operators},\label{eq:exp:hqet}
\end{align}
where $\widetilde{K}(\alpha,\beta, m, t, \widetilde{\mu};\alpha_s)$ can also be expanded in series of $\alpha_s$:
\begin{align}
\widetilde{K}(\alpha,\beta,& m, t, \widetilde{\mu};\alpha_s)=\widetilde{K}^{(0)}(\alpha,\beta)\nonumber\\
&+\frac{\alpha_s C_F}{2\pi}\widetilde{K}^{(1)}(\alpha,\beta, m, t, \widetilde{\mu})+\mathcal{O}(\alpha_s^2), 
\end{align}
with $\widetilde{K}_0(\alpha,\beta)=\delta(\alpha)\delta(\beta)$. Our result for the one-loop term is
\begin{align}
&~~~~\widetilde{K}^{(1)}(\alpha,\beta, m, t, \widetilde{\mu})\nonumber\\
=&-\bigg[\ln^2(i  t e^{\gamma_E} \widetilde{\mu})+\frac{5\pi^2}{24}\bigg]\delta(\alpha)\delta(\beta)\nonumber\\
&-\bigg[\ln(i t e^{\gamma_E} m )-\frac14\ln\frac{\widetilde\mu^2}{m^2}+2\bigg]\delta(\alpha)\delta(\beta)\nonumber\\
&+\bigg[\frac{\bar \alpha}{\alpha}\ln\frac{\widetilde\mu^2}{\alpha^2 m^2}-\ln(i t e^{\gamma_E} \alpha m  )-\frac{2}{\alpha}\bigg]_+\delta(\beta).\label{eq:kernel:hqet}
\end{align}
The first term arises from the interaction between the heavy quark and Wilson line.
A similar result in which the  collinear divergence is regularized in DR was reported in Refs.~\cite{Kawamura:2008vq,Kawamura:2018gqz}.  In Eq.~\eqref{eq:kernel:hqet} the $1/\epsilon$ and $1/\epsilon^2$ poles have already been subtracted in $\overline{\mathrm{MS}}$. We note that to reproduce the LCDA in Ref.~\cite{Bell:2008er} one should perform the Fourier transform before subtracting the $1/\epsilon^i$ poles, during to the $\ln i t$ singularities.  In contrast to QCD, the HQET nonlocal operator is non-analytic when $z \to 0$ because of the logarithmic and double-logarithmic dependence on $t$, therefore can not approach to local operator smoothly, and the local OPE does not exist~\cite{Braun:2003wx}.  The singularities at $z\to 0$ also lead to the $1/\omega$ behavior in $B$-meson LCDA $\phi_B^+(\omega)$ at $\omega\to \infty$.


With the one-loop corrections to QCD and HQET operators, we are now able to see how factorization formula Eq.~\eqref{eq:fac} works. Since the matching coefficient is calculable in perturbation theory, one can expand it in series of $\alpha_s$
\begin{align}
C(\alpha,\beta,& t ,M,\mu,\widetilde{\mu};\alpha_s)
=C^{(0)}(\alpha,\beta)\nonumber\\
&+\frac{\alpha_s C_F}{2\pi} C^{(1)}(\alpha,\beta, t ,M,\mu,\widetilde{\mu})+\mathcal{O}(\alpha_s^2).
\end{align} 
At tree-level, the QCD and HQET operators are same, so the factorization formula Eq.~\eqref{eq:fac} holds and the tree-level matching coefficient is simply $C_0(\alpha,\beta)=\delta(\alpha)\delta(\beta)$.

The one-loop  matching coefficient can be extracted by comparing the $\mathcal{O}(\alpha_s)$ terms on the both sides of Eq.~\eqref{eq:fac}, the result is
\begin{align}
C^{(1)}(\alpha,\beta, t ,M,\mu,\widetilde{\mu})=&K^{(1)}(\alpha,\beta, m, M, \mu)e^{-i M\beta t}\nonumber\\
&-\widetilde{K}^{(1)}(\alpha,\beta, m, t, \widetilde{\mu}).\label{eq:diff}
\end{align}
The reason for the phase factor $e^{-i M\beta t}$ is following:
the radiative correction changes the location of heavy quark in QCD operator from $0$ to $\beta z$, then according to Eq.~\eqref{eq:exp:field}, the heavy quark in QCD and HQET is related by a phase factor $e^{-i M \beta t}$ at leading order of $1/M$ expansion, this phase factor finally enters the matching coefficient. In momentum representation, it turns the residue momentum of heavy-quark to the total momentum.

By recalling Eqs.~\eqref{eq:kernel:qcd}, \eqref{eq:kernel:hqet} and \eqref{eq:diff}, one can evaluate the matching coefficient at one-loop level, the value reads
\begin{widetext}
\begin{align}
&C^{(1)}(\alpha,\beta, t,M,\mu,\widetilde{\mu})
=\delta(\beta)\bigg[\ln( i t e^{\gamma_E} \bar \alpha \mu  )+\frac{\bar \alpha}{\alpha}\ln\frac{\mu^2}{\widetilde{\mu}^2}+\frac32\bigg]_+ +\delta(\alpha)\bigg[\frac{\bar \beta}{\beta}\ln\frac{\mu^2}{\beta^2 M^2}\bigg]_+  e^{-i\beta M t} \nonumber\\
&~~~~+\delta(\alpha)\delta(\beta)\bigg[\ln^2(i t e^{\gamma_E}\widetilde{\mu} )+ \ln(i t e^{\gamma_E} M  )-\frac14\ln\frac{\widetilde{\mu}^2}{M^2}+\frac{5\pi^2}{24}-1\bigg]+\bigg[\frac{\bar \beta}{\beta}+\ln\frac{\mu^2}{\beta^2 M^2}\bigg]_+ e^{-i\beta M t} .\label{eq:matchingrelation}
\end{align}
\end{widetext}

One can see that the collinear divergences in QCD and HQET operators, which are represented by $\ln m^2$, are canceled. The matching coefficient $C^{(1)}(\alpha,\beta,t,M,\mu,\widetilde{\mu})$ is free of collinear and soft singularities, indicating that the factorization also holds at one-loop level. By sandwiching the both sides of matching equation between vacuum and meson sates, then performing Fourier transforms that demanded by the definitions of LCDAs, one can get the matching formula for $B$-meson LCDAs defined in QCD and HQET, which has been addressed in Ref.~\cite{Ishaq:2019dst}. However, the full result for QCD operator, Eq.~\eqref{eq:coordinate}, can not be matched onto HQET, because the $\ln m^2$ terms in Eqs.~\eqref{eq:coordinate} and \eqref{eq:kernel:hqet} do not match. This indicates that the factorization only holds at the leading power of $1/M$ expansion. 

We also note that only the $\slashed n\gamma^5$ component of axial-current is considered in this paper. If the analysis is performed for all the components, i.e., $\gamma^{\mu}\gamma^5$, Lorentz structures like $z^{\mu}\gamma^5$ and many others will enter the expansion formula. HQE for a general current will be a straightforward generalization of this work.


In summary, we have generalized the heavy quark expansion to nonlocal heavy-light current on the light-cone. Based on a perturbative calculation in operator form, we confirm up to one-loop accuracy that the QCD nonlocal heavy-light current can be matched onto the corresponding HQET operator by a factorization theorem. All soft singularities are canceled, both for QCD and HQET operators; while the collinear singularities are common and can be canceled between QCD and HQET operators. The matching coefficient is determined at one-loop and leading power of $1/M$ expansion, which does not involve any infrared scale. The matching between leading twist LCDAs defined in QCD and HQET can be derived by taking matrix elements and Fourier transforms. The results presented in this paper might be useful to resum the large logarithms of $ Q/M$ and $ M/\Lambda_{\mathrm{QCD}}$. Furthermore, if the $B$-meson LCDA in QCD is calculable by lattice QCD through large momentum effective theory, it would provide another way of accessing $B$-meson LCDA in HQET comparing with Ref.~\cite{Wang:2019msf}.

The work reported in this paper can be generalized along many directions:
(a)  It will be straightforward of applying the method described in this paper to study other  heavy-light currents on the light-cone; (b)
It will be also interesting to study the heavy quark expansion for nonlocal heavy-heavy operators; (c) In this paper the nonlocal current is located on light-cone. A study on the heavy quark expansion for equal-time operators would be important for lattice simulations of heavy meson LCDAs, through large momentum effective theory or Ioffe time pseudo-distribution approach.

\section*{Acknowledgments}
I am grateful to 
Anatoly Radyushkin 
for the discussions on coordinate representation, and to Ji Xu for sharing the results on LCDAs of heavy-light mesons. I also thank Jian-Wei Qiu, Anatoly Radyushkin, Wei Wang, Yu-Ming Wang and De-Shan Yang for reading the manuscript and valuable comments. This work is supported by Jefferson Science Associates, LLC under  U.S. DOE Contract \#DE-AC05-06OR23177 and by U.S. DOE Grant \#DE-FG02-97ER41028.


\begin{thebibliography}{widestlabel}

\bibitem{Georgi:1990um} 
H.~Georgi,
Phys.\ Lett.\ B {\bf 240}, 447 (1990).
doi:10.1016/0370-2693(90)91128-X


\bibitem{Eichten:1989zv} 
E.~Eichten and B.~R.~Hill,
Phys.\ Lett.\ B {\bf 234}, 511 (1990).
doi:10.1016/0370-2693(90)92049-O


\bibitem{Isgur:1989vq} 
N.~Isgur and M.~B.~Wise,
Phys.\ Lett.\ B {\bf 232}, 113 (1989).
doi:10.1016/0370-2693(89)90566-2


\bibitem{Neubert:1993mb} 
M.~Neubert,
Phys.\ Rept.\  {\bf 245}, 259 (1994)
doi:10.1016/0370-1573(94)90091-4
[hep-ph/9306320].


\bibitem{Manohar:2000dt} 
A.~V.~Manohar and M.~B.~Wise,
Camb.\ Monogr.\ Part.\ Phys.\ Nucl.\ Phys.\ Cosmol.\  {\bf 10}, 1 (2000).


\bibitem{Beneke:1999br} 
M.~Beneke, G.~Buchalla, M.~Neubert and C.~T.~Sachrajda,
Phys.\ Rev.\ Lett.\  {\bf 83}, 1914 (1999)
doi:10.1103/PhysRevLett.83.1914
[hep-ph/9905312].


\bibitem{Beneke:2000wa} 
M.~Beneke and T.~Feldmann,
Nucl.\ Phys.\ B {\bf 592}, 3 (2001)
doi:10.1016/S0550-3213(00)00585-X
[hep-ph/0008255].


\bibitem{Beneke:2001at} 
M.~Beneke, T.~Feldmann and D.~Seidel,
Nucl.\ Phys.\ B {\bf 612}, 25 (2001)
doi:10.1016/S0550-3213(01)00366-2
[hep-ph/0106067].


\bibitem{Bosch:2001gv} 
S.~W.~Bosch and G.~Buchalla,
Nucl.\ Phys.\ B {\bf 621}, 459 (2002)
doi:10.1016/S0550-3213(01)00580-6
[hep-ph/0106081].


\bibitem{Becher:2005fg} 
T.~Becher, R.~J.~Hill and M.~Neubert,
Phys.\ Rev.\ D {\bf 72}, 094017 (2005)
doi:10.1103/PhysRevD.72.094017
[hep-ph/0503263].


\bibitem{Lu:2018cfc} 
C.~D.~Lü, Y.~L.~Shen, Y.~M.~Wang and Y.~B.~Wei,
JHEP {\bf 1901}, 024 (2019)
doi:10.1007/JHEP01(2019)024
[arXiv:1810.00819 [hep-ph]].


\bibitem{Gao:2019lta} 
J.~Gao, C.~D.~Lü, Y.~L.~Shen, Y.~M.~Wang and Y.~B.~Wei,
arXiv:1907.11092 [hep-ph].


\bibitem{Grozin:1996pq} 
A.~G.~Grozin and M.~Neubert,
Phys.\ Rev.\ D {\bf 55}, 272 (1997)
doi:10.1103/PhysRevD.55.272
[hep-ph/9607366].


\bibitem{Ji:2013dva} 
X.~Ji,
Phys.\ Rev.\ Lett.\  {\bf 110}, 262002 (2013)
doi:10.1103/PhysRevLett.110.262002
[arXiv:1305.1539 [hep-ph]].


\bibitem{Ji:2014gla} 
X.~Ji,
Sci.\ China Phys.\ Mech.\ Astron.\  {\bf 57}, 1407 (2014)
doi:10.1007/s11433-014-5492-3
[arXiv:1404.6680 [hep-ph]].


\bibitem{Radyushkin:2017cyf} 
A.~V.~Radyushkin,
Phys.\ Rev.\ D {\bf 96}, no. 3, 034025 (2017)
doi:10.1103/PhysRevD.96.034025
[arXiv:1705.01488 [hep-ph]].


\bibitem{Orginos:2017kos} 
K.~Orginos, A.~Radyushkin, J.~Karpie and S.~Zafeiropoulos,
Phys.\ Rev.\ D {\bf 96}, no. 9, 094503 (2017)
doi:10.1103/PhysRevD.96.094503
[arXiv:1706.05373 [hep-ph]].


\bibitem{Ma:2014jla} 
Y.~Q.~Ma and J.~W.~Qiu,
Phys.\ Rev.\ D {\bf 98}, no. 7, 074021 (2018)
doi:10.1103/PhysRevD.98.074021
[arXiv:1404.6860 [hep-ph]].


\bibitem{Ma:2017pxb} 
Y.~Q.~Ma and J.~W.~Qiu,
Phys.\ Rev.\ Lett.\  {\bf 120}, no. 2, 022003 (2018)
doi:10.1103/PhysRevLett.120.022003
[arXiv:1709.03018 [hep-ph]].


\bibitem{Pilipp:2007sb} 
V.~Pilipp,
hep-ph/0703180.


\bibitem{Ishaq:2019dst} 
S.~Ishaq, Y.~Jia, X.~Xiong and D.~S.~Yang,
arXiv:1905.06930 [hep-ph].


\bibitem{Bell:2008er} 
G.~Bell and T.~Feldmann,
JHEP {\bf 0804}, 061 (2008)
doi:10.1088/1126-6708/2008/04/061
[arXiv:0802.2221 [hep-ph]].


\bibitem{Balitsky:1987bk} 
I.~I.~Balitsky and V.~M.~Braun,
Nucl.\ Phys.\ B {\bf 311}, 541 (1989).
doi:10.1016/0550-3213(89)90168-5


\bibitem{Altarelli:1977zs} 
G.~Altarelli and G.~Parisi,
Nucl.\ Phys.\ B {\bf 126}, 298 (1977).
doi:10.1016/0550-3213(77)90384-4


\bibitem{Dokshitzer:1977sg} 
Y.~L.~Dokshitzer,
Sov.\ Phys.\ JETP {\bf 46}, 641 (1977)
[Zh.\ Eksp.\ Teor.\ Fiz.\  {\bf 73}, 1216 (1977)].


\bibitem{Gribov:1972ri} 
V.~N.~Gribov and L.~N.~Lipatov,
Sov.\ J.\ Nucl.\ Phys.\  {\bf 15}, 438 (1972)
[Yad.\ Fiz.\  {\bf 15}, 781 (1972)].


\bibitem{Lepage:1979zb} 
G.~P.~Lepage and S.~J.~Brodsky,
Phys.\ Lett.\  {\bf 87B}, 359 (1979).
doi:10.1016/0370-2693(79)90554-9


\bibitem{Efremov:1978rn} 
A.~V.~Efremov and A.~V.~Radyushkin,
Theor.\ Math.\ Phys.\  {\bf 42}, 97 (1980)
[Teor.\ Mat.\ Fiz.\  {\bf 42}, 147 (1980)].
doi:10.1007/BF01032111


\bibitem{Efremov:1979qk} 
A.~V.~Efremov and A.~V.~Radyushkin,
Phys.\ Lett.\  {\bf 94B}, 245 (1980).
doi:10.1016/0370-2693(80)90869-2


\bibitem{Mueller:1998fv} 
D.~Müller, D.~Robaschik, B.~Geyer, F.-M.~Dittes and J.~Hořejši,
Fortsch.\ Phys.\  {\bf 42}, 101 (1994)
doi:10.1002/prop.2190420202
[hep-ph/9812448].


\bibitem{Braun:2014vba} 
V.~M.~Braun and A.~N.~Manashov,
Phys.\ Lett.\ B {\bf 734}, 137 (2014)
doi:10.1016/j.physletb.2014.05.037
[arXiv:1404.0863 [hep-ph]].


\bibitem{Braun:2016qlg} 
V.~M.~Braun, A.~N.~Manashov, S.~Moch and M.~Strohmaier,
JHEP {\bf 1603}, 142 (2016)
doi:10.1007/JHEP03(2016)142
[arXiv:1601.05937 [hep-ph]].


\bibitem{Braun:2017cih} 
V.~M.~Braun, A.~N.~Manashov, S.~Moch and M.~Strohmaier,
JHEP {\bf 1706}, 037 (2017)
doi:10.1007/JHEP06(2017)037
[arXiv:1703.09532 [hep-ph]].


\bibitem{Braun:2019qtp} 
V.~M.~Braun, A.~N.~Manashov, S.~Moch and M.~Strohmaier,
JHEP {\bf 1902}, 191 (2019)
doi:10.1007/JHEP02(2019)191
[arXiv:1901.06172 [hep-ph]].


\bibitem{Chanowitz:1979zu} 
M.~S.~Chanowitz, M.~Furman and I.~Hinchliffe,
Nucl.\ Phys.\ B {\bf 159}, 225 (1979).
doi:10.1016/0550-3213(79)90333-X


\bibitem{tHooft:1972tcz} 
G.~'t Hooft and M.~J.~G.~Veltman,
Nucl.\ Phys.\ B {\bf 44}, 189 (1972).
doi:10.1016/0550-3213(72)90279-9


\bibitem{Breitenlohner:1977hr} 
P.~Breitenlohner and D.~Maison,
Commun.\ Math.\ Phys.\  {\bf 52}, 11 (1977).
doi:10.1007/BF01609069


\bibitem{Xu:2016dgp} 
J.~Xu and D.~Yang,
JHEP {\bf 1607}, 098 (2016)
doi:10.1007/JHEP07(2016)098
[arXiv:1604.04441 [hep-ph]].


\bibitem{Wang:2017bgv} 
W.~Wang, J.~Xu, D.~Yang and S.~Zhao,
JHEP {\bf 1712}, 012 (2017)
doi:10.1007/JHEP12(2017)012
[arXiv:1706.06241 [hep-ph]].


\bibitem{Korchemsky:1987wg} 
G.~P.~Korchemsky and A.~V.~Radyushkin,
Nucl.\ Phys.\ B {\bf 283}, 342 (1987).
doi:10.1016/0550-3213(87)90277-X


\bibitem{Korchemskaya:1992je} 
I.~A.~Korchemskaya and G.~P.~Korchemsky,
Phys.\ Lett.\ B {\bf 287}, 169 (1992).
doi:10.1016/0370-2693(92)91895-G


\bibitem{Grozin:2014hna} 
A.~Grozin, J.~M.~Henn, G.~P.~Korchemsky and P.~Marquard,
Phys.\ Rev.\ Lett.\  {\bf 114}, no. 6, 062006 (2015)
doi:10.1103/PhysRevLett.114.062006
[arXiv:1409.0023 [hep-ph]].


\bibitem{Grozin:2015kna} 
A.~Grozin, J.~M.~Henn, G.~P.~Korchemsky and P.~Marquard,
JHEP {\bf 1601}, 140 (2016)
doi:10.1007/JHEP01(2016)140
[arXiv:1510.07803 [hep-ph]].


\bibitem{Kawamura:2010tj} 
H.~Kawamura and K.~Tanaka,
Phys.\ Rev.\ D {\bf 81}, 114009 (2010)
doi:10.1103/PhysRevD.81.114009
[arXiv:1002.1177 [hep-ph]].


\bibitem{Lange:2003ff} 
B.~O.~Lange and M.~Neubert,
Phys.\ Rev.\ Lett.\  {\bf 91}, 102001 (2003)
doi:10.1103/PhysRevLett.91.102001
[hep-ph/0303082].


\bibitem{Braun:2019wyx} 
V.~M.~Braun, Y.~Ji and A.~N.~Manashov,
Phys.\ Rev.\ D {\bf 100}, no. 1, 014023 (2019)
doi:10.1103/PhysRevD.100.014023, 10.3204/PUBDB-2019-02451
[arXiv:1905.04498 [hep-ph]].


\bibitem{Kawamura:2008vq} 
H.~Kawamura and K.~Tanaka,
Phys.\ Lett.\ B {\bf 673}, 201 (2009)
doi:10.1016/j.physletb.2009.02.028
[arXiv:0810.5628 [hep-ph]].


\bibitem{Kawamura:2018gqz} 
H.~Kawamura and K.~Tanaka,
PoS RADCOR {\bf 2017}, 076 (2018).
doi:10.22323/1.290.0076


\bibitem{Braun:2003wx} 
V.~M.~Braun, D.~Y.~Ivanov and G.~P.~Korchemsky,
Phys.\ Rev.\ D {\bf 69}, 034014 (2004)
doi:10.1103/PhysRevD.69.034014
[hep-ph/0309330].


\bibitem{Wang:2019msf} 
W.~Wang, Y.~M.~Wang, J.~Xu and S.~Zhao,
arXiv:1908.09933 [hep-ph].



	
\end{thebibliography}
\end{document}